\documentclass[aps,prl,twocolumn,footinbib]{revtex4-2}
\usepackage{amsmath,amsfonts,amssymb}
\usepackage{graphicx,float,calc}
\usepackage{color,bm}
\usepackage[colorlinks,urlcolor=blue,citecolor=blue,linkcolor=blue]{hyperref}
\usepackage{mathrsfs}
\usepackage[sort&compress]{natbib}

\begin{document}

\title{Invisible flat bands on a topological chiral edge}
\author{Youjiang Xu, Irakli Titvinidze, Walter Hofstetter}
\affiliation{Goethe-Universität Frankfurt, Institut für Theoretische Physik, 60438
Frankfurt am Main, Germany}

\begin{abstract}
We prove that invisible bands associated with zeros of the single-particle Green's function
exist ubiquitously at topological interfaces of 2D Chern insulators, dual to the chiral
edge/domain-wall modes. We verify this statement in a repulsive Hubbard model with
a topological flat band, using real-space dynamical mean-field theory to study
the domain walls of its ferromagnetic ground state. Moreover, our numerical
results show that the chiral modes are split into branches due to the
interaction, and that the branches are connected by invisible flat bands. Our
work provides deeper insight into interacting topological systems.
\end{abstract}
\maketitle

Flat bands are sensitive to interaction due to their vanishing band width.
For example, it is one of a few rigorous theorems regarding the Hubbard model
that, along with flat bands, even the weakest interaction can induce
a ferromagnetic ground state \cite{10.1143/PTP.99.489, PhysRevLett.99.026404,
PhysRevLett.62.1201, Mielke_1992, PhysRevLett.69.1608, LIU20191490}.
Things become even more interesting when the underlying flat bands
have non-trivial topology. Researchers have extensively studied
two-dimensional (2D) systems in which the interplay between flat Chern bands
and interaction leads to quantum Hall ferromagnets \cite{Katsura_2010,
PhysRevLett.108.046806, PhysRevB.85.085209}, non-abelian fractional Chern
insulators \cite{doi:10.1142/S021797921330017X, PhysRevX.1.021014,
PhysRevB.88.205101, Sheng2011, PhysRevLett.106.236804}, novel spin
excitations \cite{PhysRevB.104.155129, PhysRevB.99.014407} and topologically
protected chiral valley channels \cite{PhysRevB.105.064423}
. Nevertheless, new physics can emerge from this well-studied platform.

In this Letter, we report the emergence of invisible flat bands in the
domain-wall spectrum of a 2D quantum Hall ferromagnet resulting from the interplay of the Hubbard
interaction and the topological flat band. In a many-body fermionic system, an
invisible state associated with a certain characteristic frequency is a
special single-particle state being a null vector of the single-particle Green's function of
the system at that frequency. An invisible flat band is a collection of
invisible states with different momenta and the same characteristic
frequency. The existence of invisible states was first studied by Gurarie
\cite{PhysRevB.83.085426}, who pointed out that a closed and non-interacting
system doesn't possess an invisible state, and predicted that the existence
of invisible states may alter the edge-bulk correspondence of
topological insulators in the way that the topological invariant could
possibly change without closing the gap \cite{PhysRevB.83.085426}. Despite
the interesting properties of invisible states, they are not well studied.
In this Letter, we will present two important results about them. One is a proof that
invisible bands exist ubiquitously at topological interfaces of 2D Chern insulators,
which implies that they can serve as topological markers in such systems, and the other is the
aforementioned emergence of invisible flat bands.

We study the quantum Hall ferromagnet numerically by using real-space
dynamical mean-field theory (RDMFT) \cite{Snoek_2008}, which can capture topological interfaces in interacting systems \cite%
{PhysRevLett.122.010406} and also invisible states as we need.

This Letter is arranged as follows: First, we briefly review the basic
concepts about the zeros of the Green's function, and prove the existence of the
invisible bands dual to the chiral edge modes in 2D Chern insulators. Then
we introduce an interacting flat-band model, explain its ferromagnetic
nature and present its mean-field (MF) domain-wall spectrum, which is later
compared with the RDMFT results to identify the intrinsic interaction
effect, i.e, the branching of the chiral domain-wall mode and the emergence
of the invisible flat bands. Finally, we present an effective model to
describe the domain-wall physics.

\emph{Zeros of the Green's function -} In this Letter, the term Green's
function refers to the zero-temperature single-particle Green's function of a fermionic
many-body system with fixed number of particles $N$, whose formula can be
found in the Supplemental Material (SM) \cite{SM}. For a fixed frequency $\omega$, the
Green's function $\hat{G}$ is a $d$-by-$d$ matrix whose matrix element $G_{ij}$ is
a Fourier component of the propagation amplitude between two single-particle
states represented by $i$ and $j$, where $d$ is the dimension of the
single-particle Hilbert space. The zeros of the Green's function $\hat{G}$ are
defined by the roots $\omega_{m}$, $m=1,2,\dots,n_{0}$, of the equation $%
\det \hat{G}\left( \omega\right) =0$. The vanishing of the determinant means that
there is one or more null vectors of $\hat{G}\left( \omega_{m}\right) $, and we
call the single-particle states corresponding to these null vectors
\emph{invisible states} with characteristic frequency $\omega_{m}$, because they
are decoupled from the rest of the single-particle states in terms of
propagation amplitudes. In principle, the invisible states can be detected as long as $%
\hat{G}\left( \omega\right) $ can be measured, e.g., by angle-resolved
photoemission spectroscopy (ARPES) for electronic systems, or by angle-resolved
radio frequency spectroscopy for ultracold atoms \cite{Hofstetter_2018}.

Gurarie \cite{PhysRevB.83.085426} proved that the determinant takes the
following form:%
\begin{equation}
\det \hat{G}\left( \omega\right) =\frac{\prod_{m=1}^{n_{0}}\left( \omega
-\omega_{m}\right) }{\prod_{m=1}^{n_{p}}\left( \omega-\varepsilon
_{m}\right) },   \label{detG}
\end{equation}
where the poles $\varepsilon_{m}$ are the eigenenergies of the many-particle Hamiltonian
in the $\left( N + 1\right) $- and $\left( N-1\right) $-particle space. He also showed that $\omega _{m}
$ must be real numbers, and $n_{p}-n_{0}=d$. In a closed non-interacting
system, we can easily calculate the Green's function and find $n_{p}=d$,
hence $n_{0}=0$, which implies that zeros of the Green's function can only
exist in open or interacting systems. Nevertheless, for any closed system, we
can treat any of its subsystems as an open system, whose Green's function $\hat{g}$
is a submatrix of $\hat{G}$. In this way, we can obtain new zeros of $\hat{g}$,
which are not necessarily zeros of the full Green's function $\hat{G}$.
Especially, if the subsystem is chosen as a topological interface,
the emerging invisible bands associated with $\hat{g}$ can carry
information about the topology of the system, as we will discuss in the following.

\begin{figure}[tbh]
\centering
\includegraphics[width=0.45\textwidth]{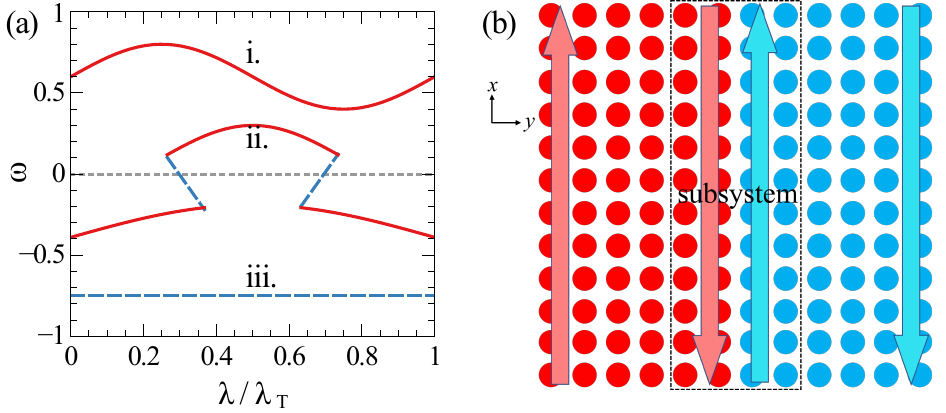}\caption{(Color online)
(a) Some possible configurations of poles and zeros as a function of $\lambda$, marked by red solid lines and blue dashed lines, respectively. The period of $\lambda$ is $\lambda_{T}$. i. A loop consisting of a single pole. ii. A loop consisting of two poles and
two zeros, forming a zig-zag path. iii. A loop consisting of a single zero. On the line $\omega=0$, we find $n_{p}^{+}=n_{p}^{-}=0$ and $n_{0}^{+}=n_{0}^{-}=1$, so Eq.~\ref{np=n0} holds. (b) Illustration of the ground state of $H$ on a $2L \times 2L$ lattice
with periodic boundary conditions filled with $L^2$ particles for each spin species. Sites half occupied by spin-up(down) particles are marked by red(blue) disks. There are two ferromagnetic domains separated by the domain walls. The arrows mark the
domain-wall spin-current channels. The dashed rectangular marks the subsystem we choose to study the domain-wall physics.
}%
\label{Fig1}%
\end{figure}

Before we show this, we need to prove an important equation describing the
configurations of the $\omega _{m}$'s and $\varepsilon _{m}$'s in an $S^{1}$
parameter space of the system. Suppose the system is parameterized by a real
number $\lambda $ so that $\det \hat{G}\left( \omega ,\lambda \right) $ is a
\emph{continuous} and \emph{periodic} function of $\lambda $. Then we
track the evolution of a pole, say $\varepsilon _{1}\left( \lambda \right) $%
, by increasing $\lambda $ from $\lambda _{0}$. Because of the periodicity
of $\det \hat{G}\left( \omega ,\lambda \right) $, $\varepsilon _{1}\left( \lambda
\right) $ must either cease to exist at some $\lambda =\lambda _{1}$, or
return to its initial value $\varepsilon _{1}\left( \lambda _{0}\right) $
after $\lambda $ is swept over its whole period one or several times,
otherwise, we will end up with infinitely many poles at any $\lambda $. In the
former case, because the difference between the number of poles and zeros $%
n_{p}-n_{0}=d$ is independent of $\lambda $, there must be one more zero
at $\lambda _{1}-0^{+}$ than at $\lambda _{1}+0^{+}$. Let's denote this zero as $%
\omega _{1}\left( \lambda \right) $. Because $\det \hat{G}\left( \omega ,\lambda
\right) $ is a continuous function of $\lambda $, the pole and the zero must
merge at $\lambda _{1}$, i.e., $\omega _{1}\left( \lambda _{1}\right)
=\varepsilon _{1}\left( \lambda _{1}\right) $. Then, we track $\omega
_{1}\left( \lambda \right) $ by decreasing $\lambda $ from $\lambda _{1}$.
Again, to avoid the existence of infinitely many zeros, $\omega _{1}\left( \lambda \right) $
ceases to exist at some point $\lambda =\lambda _{2}$, merging with either $%
\varepsilon _{1}\left( \lambda \right) $ or another pole $\varepsilon
_{2}\left( \lambda \right) $. In the latter case, we can continue tracking $%
\varepsilon _{2}\left( \lambda \right) $ to another zero $\omega _{2}\left(
\lambda \right) $ and so on, and finally we will return to $\varepsilon
_{1}\left( \lambda _{0}\right) $ and complete a loop consisting of pole- and zero-branches.
Note that every time a zero and a pole meet, the $\lambda$-direction
of tracking reverses, so the path has a zig-zag
shape. The key conclusion here is that any zero or pole is contained in such
a loop (see Fig.~\ref{Fig1}(a) for some examples). Consider now a line with fixed $%
\omega =\Omega $. The line could possibly cut through one or several such
loops and encounter poles and zeros at certain values $\Lambda _i$, $i=1,2,\dots$, i.e.,
$f\left( \lambda = \Lambda _i \right) =\Omega $ where $f$ represents either a pole $\varepsilon _m$ or a zero $\omega _m$. We say it is a positive encounter at $\Lambda _i$ if $
f\left( \Lambda _i +0^{+}\right) > \Omega > f\left( \Lambda _i -0^{+}\right) $,
and a negative encounter if $
f\left( \Lambda _i +0^{+}\right) < \Omega < f\left( \Lambda _i -0^{+}\right) $. Due to the loop structure, the
following equation holds \cite{SM}:%
\begin{equation}
n_{p}^{\left( +\right) }-n_{p}^{\left( -\right) }=n_{0}^{\left( +\right)
}-n_{0}^{\left( -\right) },  \label{np=n0}
\end{equation}%
where $n_{p}^{\left( \pm \right) }$ and $n_{0}^{\left( \pm \right) }$ denote
the number of positively (negatively) encountered poles and zeros,
respectively. This equation describes a topological property of $\det \hat{G}\left(
\omega ,\lambda \right) $ which holds for any system as long as $\lambda $
lives in a $S^{1}$ space.

Now, apply Eq.~(\ref{np=n0}) to a 2D Chern insulator. Suppose the system is periodic along the $x$ direction
and has a topological interface in the $y$ direction, e.g., an open
boundary. In such a system, we can parameterize
the Green's function $\hat{G}\left( \omega,k_{x}\right) $ by $k_{x}$, the momentum
in the $x$ direction. Instead of the whole system, we now focus on the subsystem
defined as a stripe covering the topological interface. This subsystem has the
Green's function $\hat{g}\left( \omega,k_{x}\right) $. Suppose the chemical potential sits in a band gap. If the
chemical potential is crossed by in total $C$ edge modes with the same chirality, which contribute $%
C$ poles to $\det \hat{g}\left( \omega,k_{x}\right) $, Eq.~(\ref{np=n0}) implies
that there will also be $C$ invisible bands of $\hat{g}\left( \omega,k_{x}\right) $
crossing the chemical potential with the same chirality, which establishes a one-to-one
duality between zeros and poles at topological interfaces of 2D Chern insulators, with or
without interaction. This is the first major result of this Letter.

Moreover, in the presence of interaction, the chiral edge modes can exhibit more sophisticated structures, e.g., they can be split into branches with new invisible bands connecting these branches of poles. Surprisingly, we find
that these new invisible bands are flat in the model which we study in the following.

\emph{The model and the MF results -} We study a 2D Hubbard
model with repulsive interaction $U$:
\begin{equation}
H=\sum_{m,n,\sigma }h_{m,n}c_{m,\sigma }^{\dag }c_{n,\sigma
}+U\sum_{m}n_{m,\uparrow }n_{m,\downarrow }
\label{H}
\end{equation}%
in which $h_{m,n}\equiv t\exp \left( -\frac{\left\vert z_{m}-z_{n}\right\vert
^{2}}{2}+i\operatorname{Im}z_{m}^{\ast }z_{n}\right) $, $z_{m} \equiv \sqrt{S}\left(
x_{m}+iy_{m}\right) $, $0<S<\pi $, where $\left( x_{m},y_{m}\right) $ are the
integer coordinates of the site $m$ of a square lattice, and we set $t=1$. The non-interacting
model, which has a zero-energy topological flat band, was first discovered
by Kapit and Mueller \cite{PhysRevLett.105.215303}. The topological flat band
in the non-interacting model can be regarded as the discrete lowest Landau
level, and can be generalized to a family of topological flat bands \cite%
{PhysRevA.88.033612, PhysRevA.101.013629, PhysRevA.102.053305}. For
simplicity, we focus on the case $S=\pi /2$, in which the non-interacting
model has only a flat lower band and a dispersive upper band. We only
consider the filling ratio $1/4$. In this case, the ground state of the
interacting model with $U>0$ is fully ferromagnetic, i.e., all the spins are
polarized and the flat band for the polarized spin will be fully occupied.
The energy of such a state is zero because the flat band is at zero-energy
and there is no interaction energy because of the polarization. A
zero-energy state must be the ground state of $H$ because $H$ is positive
semidefinite. The other ground states of $H$ can be obtained by the spin $%
SU\left( 2\right) $ symmetry.

Because the numbers of spin-up and spin-down particles are conserved
quantities, we can fix them and let them be equal. Now, the ground state is
no longer homogeneous. The lattice is split into two ferromagnetic domains
to lower the energy, and the ground state seeks the shortest domain wall.
Consider a $2L\times 2L$ lattice with periodic boundary conditions in both directions. Then,
there are $L^{2}$ particles for each spin species, and the domain walls will
be formed as shown in Fig.~\ref{Fig1}(b).

\begin{figure*}[tbh]
\centering
\includegraphics[width=\textwidth]{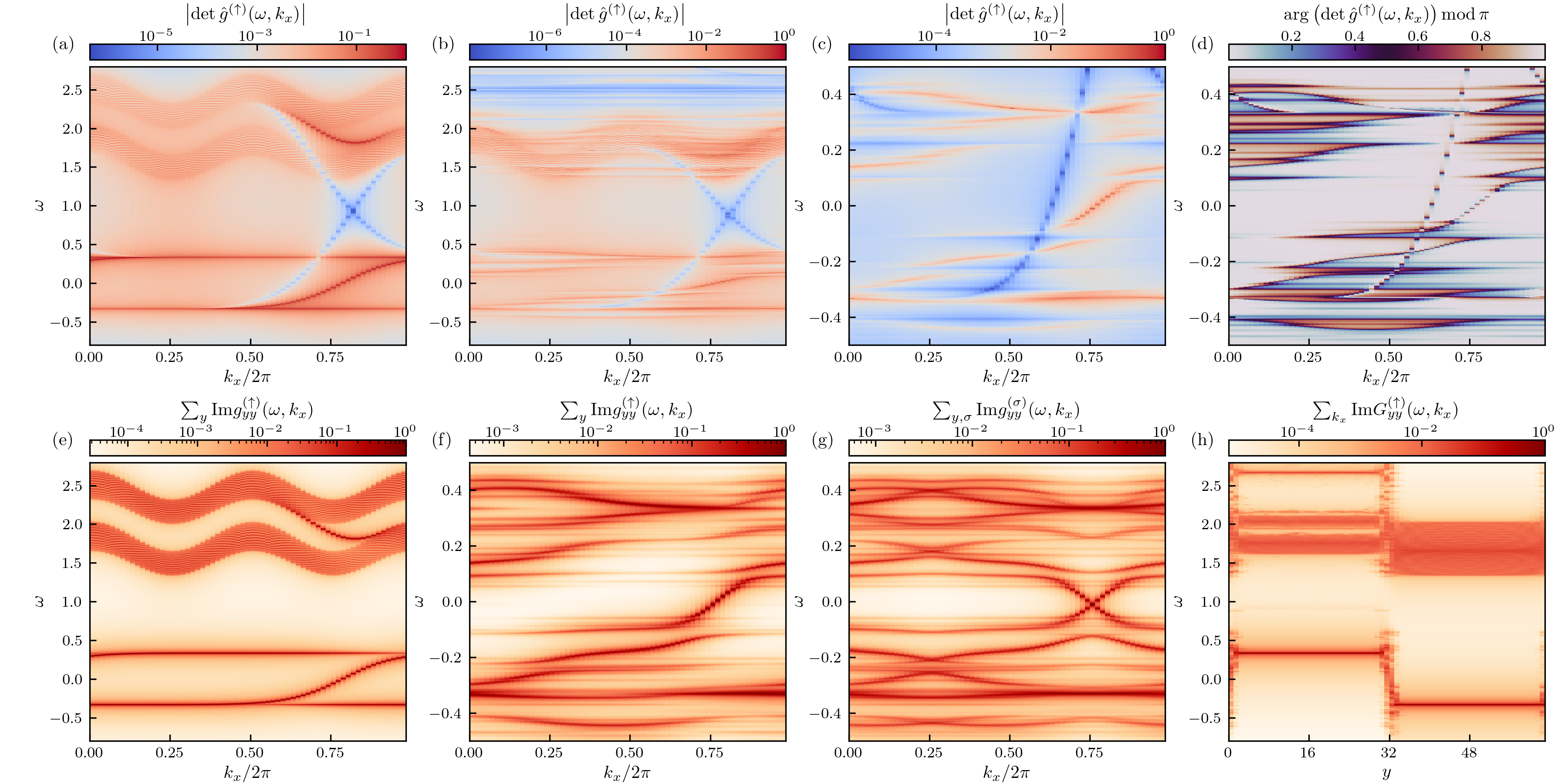}
\caption{(Color online)
The Green's functions $\hat{g}(\omega, k_x)$ (subsystem) and $\hat{G}(\omega, k_x)$ (full system) calculated by (a)(e) MF theory and (b-d)(f-h) RDMFT. (a-d) plot $\det \hat{g}$ to show the invisible bands (the darkest blue lines in (a-c)) and (e-h) plot the
$y$-summed spectral densities to show the poles (the darkest red lines), including the chiral domain-wall modes. The system size is 64-by-64, and $U$ is set as $1.33$ for MF and $2$ for RDMFT such that they produce the same gap $\Delta = 0.67$. There are
two $\omega$-ranges. The larger range includes all the bulk bands in MF results, and the smaller range resolves the details of domain-wall states. 
}%
\label{Fig2}%
\end{figure*}

Consider the MF theory for the spin-up single-particle excitations of the
ground state. The MF Hamiltonian is obtained by treating the interaction
term $U\sum_{m}n_{m,\uparrow }n_{m,\downarrow }$ as a background potential $%
U\sum_{m}n_{m,\uparrow }\left\langle n_{m,\downarrow }\right\rangle $, which
removes the entanglement between the two spin species. The MF theory gives
the correct ground-state energy only when $U$ is small compared to the gap
between the two bands in the non-interacting model. Deep in the spin-up
domain, the background potential vanishes because no spin-down particle
is present, and deep in the spin-down domain, the potential becomes $%
U/2$ because each site holds on average half of a
spin-down particle. As a result, apart from the contribution from the domain
walls, the MF spectrum is obtained by duplicating the spectrum of the
non-interacting model and shifting the duplicate by $\Delta =U/2$, the gap
between the two flat bands in the different domains. The topological bands will
contribute chiral modes to the domain wall, and the domain wall becomes a
spin current channel (see Fig.~\ref{Fig1}(b)), which is called quantum valley Hall
effect in some works \cite{doi:10.1063/1.4803084, PhysRevB.88.161406,
Ma_2016,PhysRevX.9.031021}. To visualize the domain-wall modes and their
dual invisible bands, we exactly diagonalize the MF Hamiltonian on a $%
64\times 64$ lattice with periodic boundary conditions, choose the subsystem
as a 4-site-width stripe over the domain wall, and calculate
the determinant $\det \hat{g}^{\left( \uparrow \right) }\left( \omega
+0^{+}i,k_{x}\right) $ (Fig.~\ref{Fig2}(a)) and the domain-wall spin-up spectral function $%
\sum_{y}\operatorname{Im}g_{yy}^{\left( \uparrow \right) }\left( \omega
+0^{+}i,k_{x}\right) $ (Fig.~\ref{Fig2}(e)), where the summation only takes values of $y$ within the
subsystem. We see that cross-shaped invisible bands appear, which makes Eq.~(\ref{np=n0}) hold.

\emph{RDMFT analysis on interacting system - }RDMFT is an extension of the
single-site DMFT to deal with inhomogeneity in a correlated lattice system. It solves
the lattice many-body problem by reducing the full problem to a set of
single impurity problems, one for each lattice site, with the impurity Green's function
and self-energy being the same as those of the respective site. These impurity
problems are coupled via the lattice Dyson equation,
\begin{equation}
\left[ \hat{G}^{\left( \sigma\right) }\left( \omega\right) \right]
_{m,n}^{-1}=\omega+\mu-h_{m,n}-\delta_{m,n}\Sigma_{n}^{\left( \sigma\right)
}\left(\omega\right),
\end{equation}
where $\mu$ is the chemical potential, $h_{m,n}$ is defined below Eq.~(\ref{H}), and $\Sigma_{n}^{\left( \sigma\right) }\left(\omega\right)$ is the site-dependent local
self-energy,
while the non-local parts of the self-energy are neglected in the approximation. We apply
RDMFT to a lattice with $64\times64$ sites and periodic boundary
conditions. Because RDMFT is formulated in the grand canonical ensemble, the number
of spin-up or spin-down particles cannot be fixed. In order to form two
ferromagnetic domains and keep the numbers of spin-up and spin-down
particles equal, we initiate the RDMFT iterations with a slightly biased
chemical potential to attract the spin-up(down) particles to the left(right)
half of the lattice, and the bias is removed after the first iteration. This
initial condition leads to a well converged solution from which we read that
the magnetization $\left\vert \left\langle n_{n,\uparrow}-n_{n,\downarrow
}\right\rangle \right\vert $ is $\frac{1}{2}$ for site $n$ in each domain.
We also find that the spectrum does not change qualitatively when the
interaction strength $U$ varies, therefore in the following we only present
the results for $U=2$. The inverse temperature is set as $\beta=200$, which
makes the results effectively characterize zero-temperature physics.

In Fig.~\ref{Fig2}(h), the $x$-momentum-integrated spin-up spectral density $%
\sum_{k_{x}}\operatorname{Im}G_{yy}^{\left( \uparrow \right) }\left( \omega
+0^{+}i,k_{x}\right) $ calculated by RDMFT clearly distinguishes the two
domains. The flat bands survive the interaction, however, the gap between
them $\Delta =0.67$ is much smaller than the MF value $U/2=1$. The chemical
potential is chosen as half the gap, $\mu =\Delta /2=0.33$. As a result, in
the spin-up domain, we have poles located at $\omega =-\mu $ corresponding
to creating holes in the filled spin-up flat band. And, in the spin-down
domain, we have poles located at $\omega =\mu $ corresponding to creating
spin-up particles in the filled spin-down flat band. Further numerical
results show that $\Delta $ is a monotonous function of $U$, $\Delta
\rightarrow U/2$ when $U\rightarrow 0$, and $\Delta <\Delta _{\mathrm{\max }%
}\approx 1.4$ when $U\rightarrow \infty $~\cite{SM}, which means that $\Delta $ is
always smaller than the band gap of the non-interacting model, preventing a
qualitative change of the spectrum. $\Delta _{\mathrm{\max }}$ is verified
by exact diagonalization (ED) of the Hamiltonian on a $4\times 4$ lattice,
and we find that the ED results agree surprisingly well with RDMFT in the sense
that for any $U$, the relative difference in $\Delta $ is about $1\%$~\cite{SM}.

The subsystem containing the domain wall is again chosen as a 4-site-width
stripe. We show the RDMFT determinant $\det \hat{g}\left( \omega +0^{+}i,k_{x}\right) $
in Fig.~\ref{Fig2}(b-d) and the RDMFT domain-wall spin-up spectral function $\sum_{y}%
\operatorname{Im}g_{yy}^{\left( \uparrow \right) }\left( \omega +0^{+}i,k_{x}\right) $
in Fig.~\ref{Fig2}(f). Comparing Fig.~\ref{Fig2}(e) with Fig.~\ref{Fig2}(f), we find the chiral
domain-wall mode is split into multiple branches by the interaction, and
these branches, as shown in Fig.~\ref{Fig2}(c-d), are connected by branches of
invisible bands. The zig-zag path formed by the branches can be more clearly
distinguished by the argument of $\det \hat{g}\left( \omega +0^{+}i,k_{x}\right) $ in
Fig.~\ref{Fig2}(d). Strikingly, these branches of invisible bands are flat. Note that any
linear superposition of invisible
states in the invisible flat band is an invisible state with the same
characteristic frequency. Moreover, these invisible bands, as the manifestation of quantum correlation, exist even in the full Green's function. The existence of invisible flat bands is the other
major result of this Letter.

In Fig.~\ref{Fig2}(f), if we draw a line with a fixed $\omega \in \left( -\frac{\Delta
}{2},\frac{\Delta }{2}\right) $, the line will encounter one and only one
pole, which means the bulk-edge correspondence isn't broken by the interaction. In Fig.~\ref{Fig2}(b), we see that the two
cross-shaped major invisible bands almost remain
the same as those in the MF calculation, being more stable than the chiral
modes with respect to interaction. In this sense, invisible bands may serve as a better
indicator for detecting interacting 2D Chern insulators.

Last but not least, by summing up the spin-up and the spin-down spectral
functions, we obtain Fig.~\ref{Fig2}(g) and find that the spin-up and spin-down chiral
modes connect smoothly and form bands extending over the whole Brillouin
zone. This fact inspires us to depict the domain-wall states by a 1D
single-particle model where the interaction is effectively replaced by spin
mixing, i.e., a spin-up particle can turn into a spin-down one by hopping
and vice versa. The effective model is supposed to reproduce the branching
in the domain-wall spectrum. The simplest effective model contains two bands
$\varepsilon _{\pm }\left( k\right) $ associated with eigenstates $%
\left\vert u_{k}^{+}\right\rangle =\alpha \left( k\right) \left\vert
\uparrow \right\rangle +\beta \left( k\right) \left\vert \downarrow
\right\rangle $ and $\left\vert u_{k}^{-}\right\rangle =\beta ^{*} \left( k\right) \left\vert
\uparrow \right\rangle -\alpha ^{*} \left( k\right) \left\vert \downarrow
\right\rangle $, and it is straightforward to generalize the two-band
effective model to a multi-band one to account for more branches. The
single-particle effective Hamiltonian is $h\left( k\right) =\sum\limits_{\delta=\pm
}\varepsilon _{\delta }\left( k\right) \left\vert u_{k}^{\delta
}\right\rangle \left\langle u_{k}^{\delta }\right\vert $ and the
corresponding spin-up Green's function is
\begin{equation}
g_{\mathrm{eff}}^{\left( \uparrow \right) }\left( \omega ,k\right) =\frac{%
\gamma \left( k\right) }{\omega -\varepsilon _{+}\left( k\right) }+\frac{%
1-\gamma \left( k\right) }{\omega -\varepsilon _{-}\left( k\right) },
\end{equation}%
where $\gamma \left( k\right) =\left\vert \alpha \left( k\right) \right\vert
^{2}$. The poles $\varepsilon _{\pm }\left( k\right) $ are split into
branches when $\gamma \left( k\right) =0$ or $1$ in part of the Brillouin
zone, and the invisible band connecting the poles is given by $\omega \left(
k\right) =\gamma \left( k\right) \varepsilon _{-}\left( k\right) +\left(
1-\gamma \left( k\right) \right) \varepsilon _{+}\left( k\right) $, existing
in the region where $0<\gamma \left( k\right) <1$  \cite{SM}. Apparently, it puts
a strong constraint on the Hamiltonian if we require this invisible band to
be flat.

\emph{Conclusion - }Using RDMFT, we numerically study the ferromagnetic
ground state and its domain walls of a repulsive Hubbard model, in which the
ferromagnetism comes from the topological flat band of the underlying
non-interacting model. Our results show that the flat bands survive the
interaction and they contribute chiral modes to the domain walls because of
their non-trivial topology. The chiral modes are associated with the poles
of the Green's function of the domain wall $\hat{g}$. Moreover, $\hat{g}$ also
possesses zeros associated with invisible bands. We analyse the
configurations of zeros and poles of Green's functions, and rigorously prove
that there is a one-to-one duality between the chiral modes and invisible
bands at topological interfaces of 2D Chern insulators. Therefore, the invisible bands can
also serve as indicators of non-trivial topology. Our numerical results
agree with this analysis, clearly showing the invisible bands dual to the domain-wall
modes. Moreover, the RDMFT results show that the domain-wall modes are split
into branches, an effect which is not present in the MF results, suggesting that the
branching is intrinsically caused by quantum correlations. The branching can be
captured by a 1D effective model with spin-mixing. Both the numerical and
analytical study show that the branches of domain-wall modes are connected
by invisible bands, and RDMFT shows that these invisible bands are flat. The
flatness of these invisible bands puts strong constraints on the Hamiltonian,
and could be closely related to the flat bands in the non-interacting
model.

\begin{acknowledgments}
We thank Arijit Dutta for discussion. This work was supported by the
Deutsche Forschungsgemeinschaft (DFG, German Research Foundation) under
Project No. 277974659 via Research Unit FOR 2414. This work was also
supported by the DFG via the high performance computing center \textit{%
Center for Scientific Computing} (CSC).
\end{acknowledgments}


\end{document}


\title{Supplemental material: Invisible flat bands on a topological chiral edge}
\author{Youjiang Xu, Irakli Titvinidze, Walter Hofstetter}
\affiliation{Institut für Theoretische Physik, Goethe-Universität, 60438
Frankfurt am Main, Germany}

\maketitle

\section{Green's function and its determinant}
In a zero-temperature fermionic many-body system which conserves the number of
particles, the time-ordered single-particle Green's function is given by the following
formula:%
\begin{equation}
G_{ij}\left(  t\right)  =-i\left\langle N,0|Tc_{i}\left( t\right)
c_{j}^{\dag}|N,0\right\rangle ,
\end{equation}
where $c_{i}\left(  t\right)  $ is the fermionic annihilator for state $i$ at
time $t$, $T$ is the time ordering operator, and $\left\vert N,0\right\rangle
$ denotes the ground state with $N$ particles. $N$ can be adjusted by setting
the chemical potential included in the Hamiltonian. And, by adding a constant
to the Hamiltonian, we can shift the ground state energy to zero, and all the
other eigenenergies of the Hamiltonian will be non-negative. Under these
conditions, the Green's function in the frequency domain used in the Letter is
given by%
\begin{align}
G_{ij}\left(  \omega\right) =&\int_{-\infty}^{\infty}e^{i\omega t}%
G_{ij}\left(  t\right)  dt\nonumber\\
  =&\sum_{n}\frac{\left\langle N,0\right\vert c_{i}\left\vert
N+1,n\right\rangle \left\langle N+1,n\right\vert c_{j}^{\dag}\left\vert
N,0\right\rangle }{\omega-E_{N+1,n}+0^{+}i}\\
&+\sum_{n}\frac{\left\langle
N,0\right\vert c_{j}^{\dag}\left\vert N-1,n\right\rangle \left\langle
N-1,n\right\vert c_{i}\left\vert N,0\right\rangle }{\omega+E_{N-1,n}-0^{+}i},
\end{align}
where $\left\vert N\pm1,n\right\rangle $ denotes the eigenstates of the
Hamiltonian with one more or less particles than the ground state. This formula
can be written in a compact form:%
\begin{equation}
G_{ij}\left(  \omega\right)  =\sum_{m}\frac{U_{m,i}U_{m,j}^{\ast}}%
{\omega-\varepsilon_{m}},
\end{equation}
\qquad where $\varepsilon_{m}$ can either represent $E_{N+1,n}$ or
$-E_{N-1,n}$, and $U_{m,i}$ can either represent $\left\langle N,0\right\vert
c_{i}\left\vert N+1,n\right\rangle $ or $\left\langle N-1,n\right\vert
c_{i}\left\vert N,0\right\rangle $. The completeness of the states $\left\vert
N\pm1,n\right\rangle $ in their respective particle number spaces implies that $U_{m,i}$ satisfies the following equation:%
\begin{equation}
\sum_{m}U_{m,i}U_{m,j}^{\ast}=\delta_{ij}. \label{complete}%
\end{equation}

Now consider the form of $\det \hat{G}\left(  \omega\right)  $. We denote the
dimension of the single-particle Hilbert space as $d$, which also equals the
number of orthonormal single-particle states $i$ corresponding to the operators $c_i$. In a non-interacting closed system,
there exists a set of $c_{i}$ such that each $i$ corresponds to one eigenstate of
the single-particle Hamiltonian and $U_{m,i}=\delta_{m,i}$. In this case, $\det \hat{G}\left(  \omega\right)  =\prod_{m=1}^{n_{p}}\left(
\omega-\varepsilon_{m}\right)  ^{-1}$ where $n_{p} = d$ is the number of poles in
$\det \hat{G}\left(  \omega\right)  $. For general
$c_{i}$'s, $U_{m,i}$ will be a unitary matrix, and $\det \hat{G}\left(
\omega\right)  $ is invariant under transformation of basis. Therefore, $\det
\hat{G}\left(  \omega\right)  $ has no zeros for a non-interacting closed system.

For an interacting or open system, zeros can show up. The asymptotic behavior
of the determinant is given by $\det \hat{G}\left(  \omega\right)  \rightarrow
\omega^{-d}$ when $\omega\rightarrow\infty$, so $\det \hat{G}\left(  \omega\right)
$ can no longer be $\prod_{m=1}^{n_{p}}\left(  \omega-\varepsilon_{m}\right)
^{-1}$, which drops faster than $\omega^{-d}$ in an interacting or open system
because $n_{p}>d$. Intuitively, we expect $\prod_{m=1}^{n_{p}}\left(
\omega-\varepsilon_{m}\right)  ^{-1}$ should be multiplied by a
degree-$\left(  n_{p}-d\right)  $ polynomial $\prod_{m=1}^{n_{p}-d}\left(
\omega-\omega_{m}\right)  $ to reproduce the correct asymptotic behavior,
where $\omega_{m}$'s are some complex numbers. Indeed, it was rigorously
proved that \cite{PhysRevB.83.085426}
\begin{equation}
\det \hat{G}\left(  \omega\right)  =\frac{\prod_{m=1}^{n_{p}-d}\left(  \omega
-\omega_{m}\right)  }{\prod_{m=1}^{n_{p}}\left(  \omega-\varepsilon
_{m}\right)  }, \label{detG}%
\end{equation}
where $\omega_{m}$ are real numbers.

\section{Proof of Eq.~(2)\ in the main text}
Consider a loop composed of branches of zeros $\omega_{m}\left(
\lambda\right)  $ and poles $\varepsilon_{m}\left(  \lambda\right)  $ in
the $S^{1}$-space of $\lambda$, where $m=1,2,\dots,n_b$. It is straightforward to
generalize the proof to the case in which the loop consists of a single zero
or pole. Suppose $\varepsilon_{m}\left(  \lambda\right)  $ merges with
$\omega_{m}\left(  \lambda\right)  $ at $\lambda_{2m-1}$, $\omega_{m}\left(
\lambda\right)  $ merges with $\varepsilon_{m+1}\left(  \lambda\right)  $ at
$\lambda_{2m}$, where $\varepsilon_{n_b + 1}\left(  \lambda\right)  $ is defined as
$\varepsilon_{1}\left(  \lambda\right)  $. Without loss of generality, we
assume $\lambda_{1}>\lambda_0 \equiv \lambda_{2n_b}$, and we parameterize the loop by $\tau
\in\left[  0,1\right)  $ in the following way:%
\begin{equation}
\tilde{\lambda}\left(  \tau\right) \equiv \left(  \lambda_{m}-\lambda
_{m-1}\right)  \left(  2 n_b \tau - m\right) +\lambda_{m},~\tau\in\left[
\frac{m-1}{2n_b},\frac{m}{2n_b}\right),~m=1,2,\dots,2n_b,
\end{equation}
and
\begin{equation}
\tilde{\omega}\left(  \tau\right) \equiv \left\{
\begin{array}
[c]{c}%
\varepsilon_{m}\left(  \tilde{\lambda}\left(  \tau\right)  \right)  ,~\tau\in\left[  \frac{2m-2}{2n_b},\frac{2m-1}{2n_b}\right) \\
\omega_{m}\left(  \tilde{\lambda}\left(  \tau\right)  \right)  ,~\tau\in\left[  \frac{2m-1}{2n_b},\frac{2m}{2n_b}\right)
\end{array}
\right.  ,~m=1,2,\dots,n_b.
\end{equation}

Now, consider a line $\omega=\Omega$. We say that $\tilde{\omega}\left(
\tau\right)$ positively encounters the line at $T$ if $\tilde{\omega
}\left(  T+0^{+}\right)  >\tilde{\omega}\left(  T\right)
=\Omega>\tilde{\omega}\left(  T-0^{+}\right)  $, and that $\tilde{\omega}\left(
\tau\right)  $ negatively encounters the line at $T$ if $\tilde{\omega
}\left(  T+0^{+}\right)  <\tilde{\omega}\left(  T\right)
=\Omega<\tilde{\omega}\left(  T-0^{+}\right)  $. Similarly, we can define
the positive or negative encounter of a zero or pole with the line
$\omega=\Omega$.\ Let $f\left(  \lambda\right)  $ represent either a pole
$\varepsilon_{m}\left(  \lambda\right)  $ or a zero $\omega_{m}\left(
\lambda\right)  $. We say that $f\left(  \lambda\right)  $ positively encounters
the line at $\Lambda$ if $f\left(  \Lambda+0^{+}\right)  >f\left(  \Lambda\right)
=\Omega>f\left(  \Lambda-0^{+}\right)  $, and that $f\left(  \lambda\right)  $
negatively encounters the line at $\Lambda$ if $f\left(  \Lambda+0^{+}\right)
<f\left(  \Lambda\right)  =\Omega<f\left(  \Lambda-0^{+}\right)  $. As long as the
line $\omega=\Omega$ doesn't meet any merging point, i.e., $\Omega\neq
\omega_{m}\left(  \lambda_{2m-1}\right)  $ and $\Omega\neq\omega_{m}\left(
\lambda_{2m}\right)  $ for any $m$, then a positive encounter of
$\tilde{\omega}\left(  \tau\right)  $ is either a positive encounter of a
pole, or a negative encounter of a zero, because $\tilde{\lambda}'\left( \tau \right)$ is always positive for a pole and negative for a zero due to the assumption $\lambda _1 > \lambda _{0}$. Therefore, the number of positive encounters of $\tilde{\omega}\left(
\tau\right)$, $n^{\left(  +\right)  }$, equals $n_{p}^{\left(
+\right)  } + n_{0}^{\left(
-\right)  }$, where $n_{p}^{\left(
\pm\right)  }$ and $n_{0}^{\left(  \pm\right)  }$ denote the number of
positively(negatively) encountered poles and zeros, respectively. Similarly,
we have $n^{\left(  -\right)  }=n_{p}^{\left(  -\right)  }+n_{0}^{\left(
+\right)  }$. Since $\tilde{\omega}\left(  \tau\right)  $ is a loop, we have $n^{\left(  +\right)  } = n^{\left(  -\right)  }$, or:%
\begin{equation}
n_{p}^{\left(  +\right)  }+n_{0}^{\left(  -\right)  }=n_{p}^{\left(  -\right)
}+n_{0}^{\left(  +\right)  },
\end{equation}
which is identical with Eq.~(2)\ in the main text.

\section{Fit the gap calculated by RDMFT}
\begin{figure}[tbh]
\centering
\includegraphics[width=0.5\textwidth]{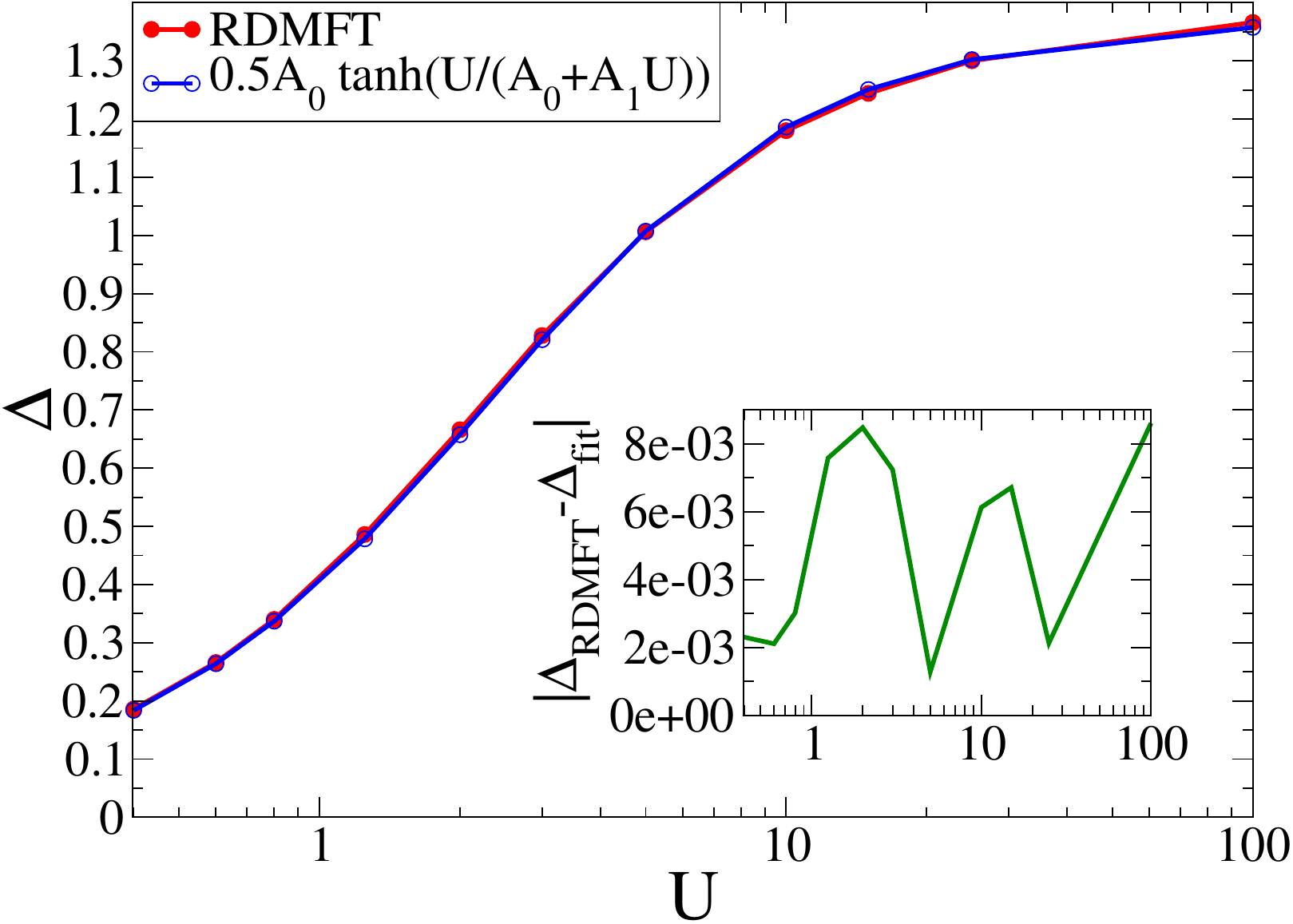}\caption{(Color online)
The gap $\Delta$ as a function of $U$. The red closed circles are the RDMFT results while the blue open circles are given by the fitting function. The difference between the two is shown in the inset panel.
}%
\label{FigSM1}%
\end{figure}

We find a simple function to fit the gap $\Delta$ calculated by RDMFT as a function of $U$:
\begin{equation}\label{fitting}
  \Delta = \frac{A_0}{2}\tanh\frac{U}{A_0 + A_1 U},
\end{equation}
in which the fitting parameters are given by $A_0 = 2.976$ and $A_1=0.619$. Note that the fitting function is approximated by $\Delta=U/2$ at small $U$, and is bounded by $\frac{A_0}{2}\tanh \frac{1}{A_1} = 1.375$ for $U \rightarrow \infty$. We compare the RDMFT results with the fitting function in Fig.~\ref{FigSM1}

\section{Comparison of the gap calculated by ED and RDMFT}
In Table.~\ref{tableSM}, we list $\Delta$, the gap between the two flat bands
in the different domains, calculated by ED and RDMFT. The results obtained by the two methods show very good agreement.

\begin{table}[ptb]
\caption{$\Delta$ calculated by ED and RDMFT}%
\label{tableSM}%
\begin{ruledtabular}
\begin{tabular}{cccc}
U & 1 & 10 & 100\\
$\Delta$ (ED) & 0.410 & 1.194 & 1.382\\
$\Delta$ (RDMFT) & 0.408 & 1.180 & 1.366\\
\end{tabular}
\end{ruledtabular}
\end{table}

\section{1D effective model for the domain wall: An example}

\begin{figure}[tbh]
\centering
\includegraphics[width=0.5\textwidth]{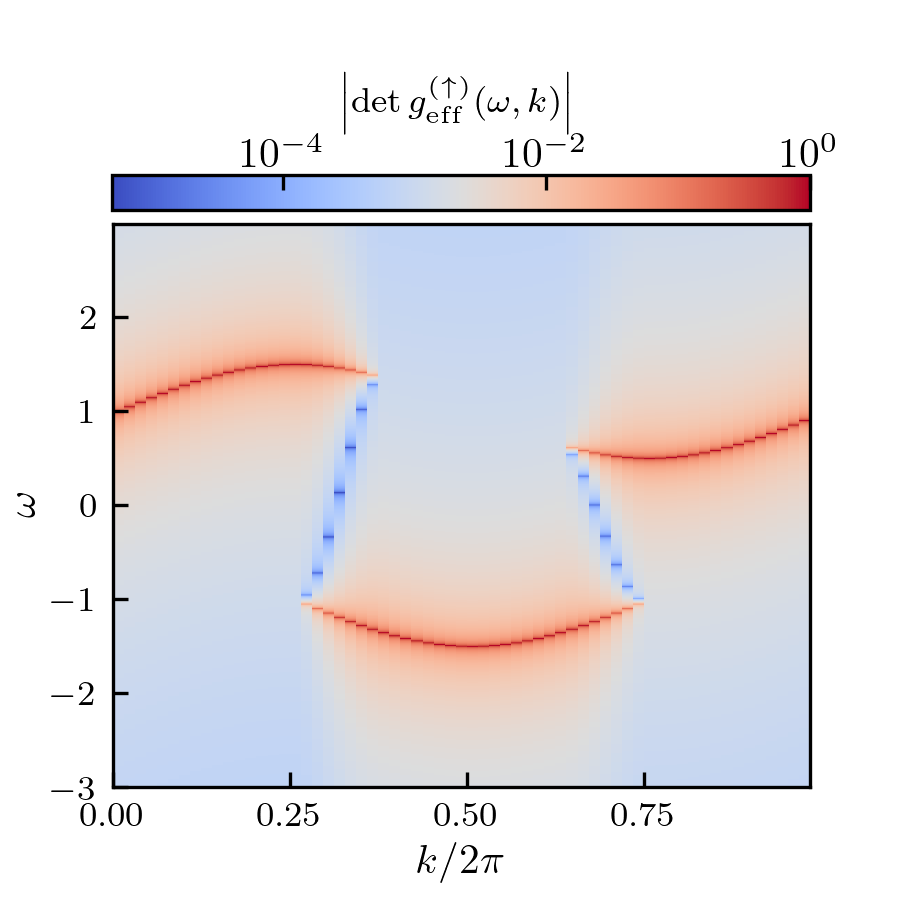}\caption{(Color online)
The branches of zeros and poles of the Green's function of the 1D effective model defined in Eq.~(\ref{ep}-\ref{beta}).
}%
\label{FigSM2}%
\end{figure}

As discussed in the main text, the branching of the chiral domain-wall mode
can be captured by a non-interacting effective model in which the interaction
is effectively replaced by spin-mixing. Here we show a two-band
model as a specific example. Suppose the two bands are given by
\begin{align}
\varepsilon_{+}\left(  k\right)    & =1+\frac{1}{2}\sin k,
\label{ep}\\
\varepsilon_{-}\left(  k\right)    & =-1+\frac{1}{2}\cos k.
\label{em}
\end{align}
The single-particle effective Hamiltonian is $h\left(  k\right)  =\sum
\limits_{\delta=\pm}\varepsilon_{\delta}\left(  k\right)  \left\vert u_{k}^{\delta
}\right\rangle \left\langle u_{k}^{\delta}\right\vert $ where the eigenstates
are given by
\begin{align}
\left\vert u_{k}^{+}\right\rangle  & =\alpha\left(  k\right)  \left\vert
\uparrow\right\rangle +\beta\left(  k\right)  \left\vert \downarrow
\right\rangle \\
\left\vert u_{k}^{-}\right\rangle  & =\beta^{\ast}\left(  k\right)  \left\vert
\uparrow\right\rangle -\alpha^{\ast}\left(  k\right)  \left\vert
\downarrow\right\rangle
\end{align}
in which
\begin{align}
\alpha\left(  k\right)    & =\theta\left(  \cos k\right)  +\theta\left(  -\cos
k\right)  \theta\left(  -\cos2k\right)  \cos2k,
\label{alpha}\\
\beta\left(  k\right)    & =\theta\left(  -\cos k\right)  \left(
\theta\left(  \cos2k\right)  +\theta\left(  -\cos2k\right)  \sin2k\right).
\label{beta}
\end{align}
where $\theta$ is the step function: $\theta\left(  x\right)  =1$ if $x\geq
0$, and $\theta\left(  x\right)  =0$ if $x<0$. We show the branches of poles
and zeros of this model in Fig.~\ref{FigSM2}.

